\documentclass[
bibnotes,
amsmath,amssymb,
aps,
twocolumn,
prb,
]{revtex4}

\usepackage{graphicx}
\usepackage{dcolumn}
\usepackage{bm}
\usepackage{color}

\def\la{~\mbox{\raisebox{-.6ex}{$\stackrel{<}{\sim}$}}~}
\def\ga{~\mbox{\raisebox{-.6ex}{$\stackrel{>}{\sim}$}}~}

\begin{document}

\title{ First principles study of electronic and structural properties of CuO}

\author{Burak Himmetoglu, Renata M. Wentzcovitch and Matteo Cococcioni}
 \affiliation{Department of Chemical Engineering and Materials Science, University of Minnesota, Minneapolis,
                Minnesota 55455}

\date{\today}

\begin{abstract}
We investigate the electronic and structural properties of CuO, which shows significant deviations from the
trends obeyed by other transition-metal monoxides. Using an extended Hubbard corrective functional, we uncover 
an orbitally ordered insulating ground state for the cubic phase of this material, which was expected but never 
found before. This insulating state results from a fine balance between the tendency of Cu to complete its 
d-shell and Hund's rule magnetism. Starting from the ground state for the cubic phase, we also study
tetragonal distortions of the unit cell (recently reported in experiments), the consequent electronic
reorganizations and identify the equilibrium structure. Our calculations reveal an unexpected richness of possible
magnetic and orbital orders, relatively close in energy to the ground state, whose stability depends on the sign
and entity of distortion.

\end{abstract}

\maketitle

\section{\label{sec:Introduction}Introduction}

Among the transition-metal oxide (TMO) compounds, CuO shows quite peculiar characteristics. At variance with other
TMOs, which crystallize in a cubic rock-salt structure (with possible rhombohedral distortions), it is found to
have a lower-symmetry monoclinic cell~\cite{kimura2008cupric,yang1989magnetic,asbrink1970refinement}. 
Similarly to other TMOs, CuO has an antiferromagnetic ground state~\cite{kimura2008cupric}. 
However, its Ne\'{e}l temperature ($T_N \simeq 220 K$) is substantially 
lower than the (expected) linear trend followed by other TMOs (The Ne\'{e}l temperatures of TMOs are observed to increase
almost linearly, from MnO ($T_N \simeq 116 K$) to NiO ($T_N \simeq 525 K$), with the nuclear charge of the 
transition metal).
The reduction in $T_N$ seems to be related to the fact that the monoclinic ground state is stabilized by a
Jahn-Teller structural distortion, which yields lower effective exchange interaction compared to a cubic 
structure~\cite{filippetti2005magnetic}. 
 
In spite of the fact that it is not stable, studying the cubic phase of this material is still 
interesting as a reference point for the characterization of all the electronic mechanisms correlating to 
the structural deformations. In addition, cubic CuO has also been recently considered as a proxy structure for high T$_c$ 
superconducting cuprates~\cite{grant2008electronic} ,to investigate the interplay between ``d" and ``p" electrons.
Although cubic CuO has never been observed experimentally, a tetragonal phase of CuO (i.e. elongated rock-salt cell
along one crystal axis) has recently been deposited on substrates of SrTiO$_3$ thin films~\cite{siemons2009tetragonal}. 
The tetragonal phase of CuO has become a subject of several theoretical studies based on density functional theory (DFT)
~\cite{grant2008electronic,peralta2009jahn,chen2009hybrid}. All the DFT studies have predicted, in agreement with
the experimental results, a distortion characterized by $1.1 \la c/a \la 1.3$
~\cite{grant2008electronic,peralta2009jahn,chen2009hybrid} (where $c$ denotes the elongated lattice parameter 
and $a$ denotes the ones in the perpendicular direction). Among possible
magnetic configurations, the antiferromagnetic-II (AF-II), characterized by ferromagnetic (111) planes with opposite spins
with respect to their neighbors,
and the AF-IV, characterized by ferromagnetic (110) planes with opposite spins with respect to their neighbors, 
configurations compete for minimum energy. 
Self-interaction corrected density functional (SIC) based study predicts an AF-II ordered ground state with 
$c/a \simeq 1.1$~\cite{peralta2009jahn}, while the hybrid density functionals predict an AF-IV ordered ground state 
with $c/a \simeq 1.3$~\cite{chen2009hybrid}.
In both studies, a local energy minimum was also identified at $c/a \simeq 0.9$. At this local minimum, the magnetic
structure was found to be AF-II. DFT+U, limited only to the AF-II magnetic ordering, yields an equilibrium structure
with $c/a \simeq 1.1$\cite{grant2008electronic}. 
In all these studies, the cubic phase (i.e. the limit when 
$c/a = 1$) is found to be metallic and corresponding to a local peak in the energy.
However, as pointed out in other studies~\cite{grant2008electronic}, 
it seems quite unlikely that the insulating
structures with $c/a < 1$ and $c/a > 1$ are "connected" by a metallic state at $c/a = 1$. Instead, an insulating
state for the cubic structure seems more reasonable.

In this paper, we revisit the cubic and tetragonal phases of CuO to investigate the underlying mechanism characterizing 
the electronic, magnetic and structural properties of this compound using a DFT+U based corrective functional within 
the AF-II magnetic order. We find an insulating ground state for the cubic phase of CuO, that was expected but never 
found before in the literature.  
Starting from this insulating ground state for the cubic cell, we also study tetragonal distortions and find an 
equilibrium structure in agreement with experiments and previous calculations. 
The properties of this ground state are controlled by an interesting interplay between Hund's rule magnetism
and electronic localization. We believe that similar effects could also play an 
important role in more complex cuprate materials. 

The paper is organized as follows: in section~\ref{sec:Method} we summarize the DFT+U method we have used.
In section~\ref{sec:cubic} we discuss the electronic structure of the cubic phase,
from DFT and DFT+U functionals. In section~\ref{sec:cubic_j}, we introduce an extension of the DFT+U
method to include an effective exchange parameter J (DFT+U+J) and discuss the resulting electronic structure
of the cubic phase. In section~\ref{sec:tetragonal} we study elongated structures and compare our 
results with those from the existing literature. Finally, in section~\ref{sec:summary} we summarize our findings 
and propose some conclusions.
%
\section{\label{sec:Method}DFT+U Method}
In this study, we employ the Hubbard model DFT+U corrective scheme, originally introduced in
~\cite{anisimov1991band,anisimov1993density,solovyev1994corrected}, 
that has become one of the most popular choices to study systems characterized by strong electronic
correlations. Although not able to capture all the possible correlated ground states, this corrective scheme 
has proved to be quite versatile in the description of the ground states of
several transition metal compounds~\cite{hsu2010cobalt,hsu2009first},
minerals of the Earth's interior~\cite{hsu2010cobalt,hsu2009first,hsu2011spin,hsu2010hubbard,stackhouse2010determination,
stackhouse2007electronic},
molecular complexes~\cite{kulik2006density,scherlis2007simulation,leung2010cobalt,kulik2010systematic},
TMOs~\cite{cococcioni2005linear,anisimov1991band,anisimov1993density,mazin1997insulating,solovyev1998orbital} 
and magnetic impurities~\cite{mattioli2010deep}. 
Other more elaborate corrective schemes have also been successfully used in the literature, 
including self-interaction corrected density functionals~\cite{filippetti2003self},
hybrid density functionals~\cite{becke1993new},
dynamical mean field theory~\cite{lichtenstein1998ab} and reduced density matrix functional 
theory~\cite{sharma2008reduced}.
Among these, DFT+U has the advantage to present low computational costs~\cite{setyawan2010high} 
and to allow for the efficient calculation of energy derivatives (e.g. forces, stresses, elastic constants etc.). 
The scheme is based on the addition of a corrective term, inspired from the Hubbard model, that favors Mott 
localization of electrons on atomic sites. The total energy functional of DFT+U can be written 
as~\cite{cococcioni2005linear}
\begin{equation}
 E_{{\rm DFT+U}} = E_{{\rm DFT}}\left[ n\left( {\bf r} \right)  \right] + 
                   E_{\rm U}\left[ \{ n^{I\, \sigma}_{m\, m'} \} \right] \label{dft_pu}
\end{equation}
where $E_{{\rm DFT}}$ is a standard approximate DFT functional and the Hubbard correction $E_U$, according to
the simplified functional by Dudarev et. al.~\cite{dudarev1998electron}, is given by
\begin{equation}
E_{\rm U} = \sum_{I, \sigma} \frac{U^I}{2}\, {\rm Tr}\left[ {\bf n}^{I\, \sigma} 
               \left( {\bf 1} - {\bf n}^{I\, \sigma} \right) \right]. \label{hub_u}
\end{equation}
In the above equation, $U^I$ is the Coulomb repulsion parameter on atomic site $I$ 
(usually applied on the d states of a transition metal) and the occupation matrices 
${\bf n}^I$ are computed as
\begin{equation}
n^{I\, \sigma}_{m\, m'} = \sum_{ {\bf k}\, v}\, f^{\sigma}_{ {\bf k}\, v}\, \langle \psi_{ {\bf k}\, v}^{\sigma} \, 
                                  \vert \phi_m^I \rangle\, \langle \phi_{m'}^I \vert 
                                    \psi_{ {\bf k}\, v}^{\sigma} \rangle \label{occup}
\end{equation}
where $\psi_{ {\bf k}\, v}^{\sigma}$ denote the Kohn-Sham states, $f^{\sigma}_{ {\bf k}\, v}$ represent their occupations 
according to the Fermi-Dirac distribution of their energy,
and $\phi_{m}^I$ are the atomic orbitals with state index $m$ and centered on site I (In this work
we use orthogonalized atomic orbitals, i.e $\langle \phi_m^{I\, \sigma} \vert \phi_m^{J\, \sigma} \rangle 
= \delta^{I\, J}$, so that orbitals centered on different atomic sites are orthogonal). 
The representation of occupation matrices in terms of atomic orbitals given in equation (\ref{occup}) is not 
the only possible choice. The same scheme can be used with different sets of wavefunctions such as Wannier 
functions~\cite{o2010projector,mazurenko2007wannier}, that may offer a more flexible representation of electronic 
localization. For the same purpose, a recent work introduced an extension to the functional of equation 
(\ref{hub_u}) to include inter-site terms~\cite{jr2010extended}. 
While we expect that the inclusion of these terms (especially those between O and Cu) might be important to refine
structural properties and to resolve some fine details in the electronic structure, in this paper we neglect them
and focus on the atomic (on-site) ones. 

In our work, the on-site Coulomb repulsion parameters $U^I$s are determined using the linear response 
approach introduced in~\cite{cococcioni2005linear}. In this work, we have generalized this approach to include 
the responses of the s states of Cu and O treated as a ``reservoir" of charge (instead of the neutralizing 
``background" of reference~\cite{cococcioni2005linear}). Our results show that inter-site interactions ($V$) are 
significantly smaller than on site ones ($U$) and our approximation is justified.   

In many cases, the DFT ground state for TMOs have different properties than the DFT+U ground state. 
For instance, DFT+U could stabilize a magnetic ground state with an insulating gap, while DFT results in a 
metallic one. 
Therefore, a more accurate determination of the $U^I$s should involve a self-consistent procedure, 
where the linear response computation is repeatedly performed on the DFT+U ground state, 
until a convergence in their values is reached\cite{kulik2006density,jr2010extended}. 
This self-consistent procedure proved to be necessary in our study due to the qualitative differences between the DFT
and the DFT+U ground states. 

In our calculations, we have used the Perdew-Burke-Ernzherof(PBE)~\cite{perdew1996generalized} GGA
functional to model the exchange-correlation energy. The Cu and O atoms are represented by ultrasoft pseudopotentials 
and the kinetic energy and charge density cut-offs are chosen to be $35$ Ry and $280$ Ry respectively. The Brillouin 
zone integrations are performed using $8\times 8\times 8$ Monkhorst and Pack special point
grids~\cite{monkhorst1976special} and a Methfessel and Paxton smearing of the Fermi-Dirac distribution
~\cite{methfessel1989high}, with a smearing width of $0.01$ Ry. 
All calculations were performed by using the
plane waves pseudopotential `pwscf' code contained in the {\it Quantum ESPRESSO} package~\cite{giannozzi2009quantum},
where we have implemented the `+J' corrections (as discussed in section~\ref{sec:cubic_j})
starting from the existing DFT+U functional.

%
\section{\label{sec:cubic}DFT and DFT+U calculations in the cubic phase}
\begin{figure}[!ht]
\includegraphics[width=0.45\textwidth]{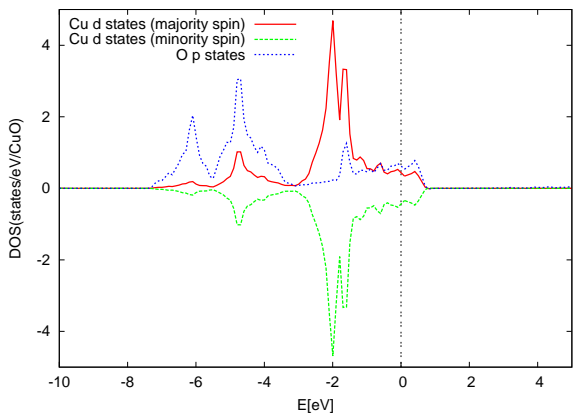}
\caption{\label{fig:cuo_gga_4} (Color online) The projected density of states calculated by the GGA
                               functional for cubic CuO.}
\end{figure}

\begin{figure}[!ht]
\includegraphics[width=0.45\textwidth]{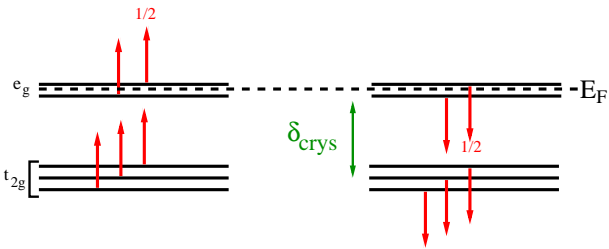}
\caption{\label{fig:d_levels_nm} (Color online) Splitting of d levels in a cubic crystal field.}
\end{figure}

\begin{figure}[!ht]
\includegraphics[width=0.45\textwidth]{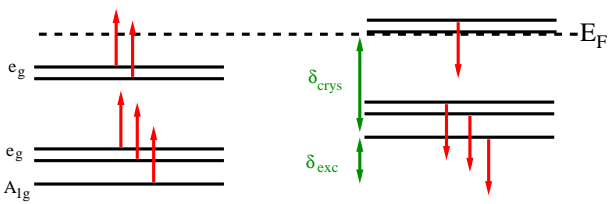}
\caption{\label{fig:d_levels} (Color online) Splitting of Cu d states in a rhombohedral field with the onset of 
                               magnetic ordering.}
\end{figure}

Previous studies of the cubic phase of CuO, based on GGA functionals, predicted a metallic and a non-magnetic ground state.
While other TMOs are also predicted to be metallic within GGA, they have an antiferromagnetic ground state with 
ferromagnetic (111) planes of transition-metal ions alternating with opposite magnetization (AF-II). This magnetic
order imposes a rhombohedral symmetry to the cell that sometimes produces a distortion. In this work, CuO is also
described with a rhombohedral cell. The unit cell consists of 4 atoms, of which the two Cu atoms have opposite spins. 
We find that the optimized structure has a lattice parameter of 4.256 \AA, which we have adopted for the rest 
of the calculations.
The density of states obtained with GGA is shown in Fig.~\ref{fig:cuo_gga_4}. As it can be observed, the GGA functional
yields a non-magnetic (due to the degeneracy between the two spin states) and metallic ground state with a finite 
contribution to density of states at the Fermi level. This result could be understood in a simple way by inspecting
the splitting of d levels of Cu in a cubic crystal field, schematically represented in Fig.~\ref{fig:d_levels_nm}.
On each Cu$^{+2}$ ion, there are 9 electrons placed in the 3d levels. The d levels are split in the cubic crystal 
field into a doubly degenerate $e_g$ (higher energy) and triply degenerate $t_{2g}$ states (lower energy). As 
illustrated in Fig.~\ref{fig:d_levels_nm}, the metallic character and the non-magnetic ground state are due to the
degeneracy of the highest energy $e_g$ states with either spin. On these 4 orbitals, Cu hosts 3 electrons, 
thus leading to partially filled bands that results in metallic ground state. It is important to notice that 
O also provides a finite contribution to the density of states at the Fermi level, thus p states (non-magnetic) 
are also partially filled. This scenario is similar to that of paramagnetic insulators, with the additional 
complication of spin degeneracy. 
The orbital degeneracy contributing to the metallic character of this ground state is obviously a consequence of the
cubic symmetry that makes the $e_g$ states equivalent. This degeneracy cannot be broken by the straight use
of DFT+U and since the Hubbard corrective functional is spin diagonal. 
\begin{figure}[!ht]
\includegraphics[width=0.45\textwidth]{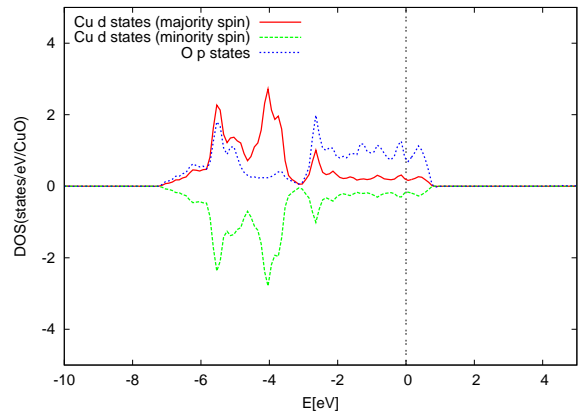}
\caption{\label{fig:cuo_gga_u} (Color online) The projected density of states calculated by the $GGA+U$ functional.
                               The on-site Hubbard parameter is $U = 9.79\, {\rm eV}$, which is calculated
                               by the linear response approach~\cite{cococcioni2005linear}.}
\end{figure}
The density of states of the ground state resulting from the GGA+U functional is shown in Fig.~\ref{fig:cuo_gga_u}
where it is evident that the main effect of the Hubbard correction consists in the (probably exaggerated) stabilization
of filled d states that shift to lower energies. Both d states ($e_g$) and  p states are left at the Fermi energy. 
Owing to the presence of O p states around the Fermi level, one might be tempted to extend the Hubbard correction to 
these states. This was indeed explored in reference~\cite{2000cond.mat..9107N}.
\begin{figure}[!ht]
\includegraphics[width=0.45\textwidth]{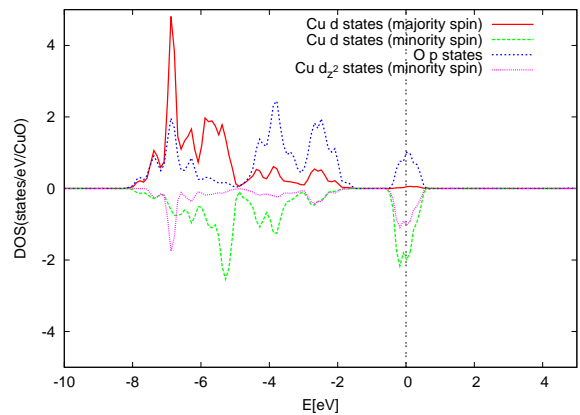}
\caption{\label{fig:cuo_gga_u_p} (Color online) The projected density of states calculated by the
                               $GGA+U+U_p$ functional. The on-site Hubbard parameters are
                               $U = 9.79\, {\rm eV}$ and $U_p = 8.47\, {\rm eV}$, which are determined by linear
                               response approach~\cite{cococcioni2005linear}.}
\end{figure}
Fig.~\ref{fig:cuo_gga_u_p} shows the density of states of CuO obtained with a Hubbard correction extended to O 
p states. The Hubbard U on O p states ($U_p$) was evaluated using the same linear response method of reference
~\cite{cococcioni2005linear}, that yielded a value of $U_p \simeq 8.47\, eV$ (vs $9.79\, eV$ of Cu). 
As evident from the density of states, while the metallic character is preserved, a magnetic ground state now emerges
from the lifting of the spin degeneracy. This new situation is schematically illustrated in Fig.~\ref{fig:d_levels},
where an exchange splitting between opposite spin levels has resulted in a magnetic ground state. 
With GGA+U, the non-magnetic ground state leads to an effective cubic symmetry (in spite of the use of
the rhombohedral unit cell), therefore the lower energy $t_{2g}$ states are degenerate. The rhombohedral symmetry,
induced by the antiferromagnetic order, lifts this degeneracy and splits them into a non-degenerate state with
$A_{1g}$ symmetry and a doublet of $e_g$ symmetry as illustrated in Fig.~\ref{fig:d_levels}.
However, the material is still metallic
due to the degeneracy of minority spin $e_g$ states.  
It is important to notice that O p states still contribute to the metallic
character (thus resulting in a partially filled p band) with equal contributions from the two spins, in spite of 
the polarization of the d states. The magnetic ground state in $GGA+U+U_p$ is not directly due to $U_p$ but, 
rather a consequence of the redistribution of electrons.
It is instructive to compare at this point, the occupations of d and p orbitals
(i.e. traces of $n^{I\, \sigma}_{m\,m'}$i given in equation (\ref{occup}) )
between the two cases (with $GGA+U$ and $GGA+U+U_p$). For $GGA+U$, we obtain
$n^{\uparrow}_{\rm Cu}(e_g) = n^{\downarrow}_{\rm Cu}(e_g) \simeq 1.84$,  
while $n_{{\rm O}_p} \simeq 4.94$. In the case of $GGA+U+U_p$ we obtain
$n^{\uparrow}_{\rm Cu}(e_g) \simeq 1.96$, 
$n^{\downarrow}_{\rm Cu}(e_g) \simeq 1.40$, 
while $n_{{\rm O}_p} \simeq 5.27$. The main consequence of using $U_p$ consists in the increase of $n_{{\rm O}_p}$
and the consequent depression of the population of the d orbitals. Thus, the magnetic ground state seems to be 
promoted by the partial (and numerically marginal) decrease in the population of d-orbitals. 
This picture is corroborated by Fig.~\ref{fig:cuo_gga_u_p}, which shows the explicit contribution to the density of 
states from $d_{z^2}$ (one of the $e_g$) states, that accounts for half of the density around the Fermi level.
It is also important to notice how the peak in the $d_{z^2}$ density of states correlate with those of the p states,
suggesting partial hybridization between Cu and O.

The emergence of the magnetic, albeit metallic ground state is due to the rhombohedral symmetry and cannot be broken
by the Hubbard corrections. Thus, the metallic character is a consequence of the crystal symmetry, similar to the 
case of FeO~\cite{cococcioni2005linear}. The effective equivalence between the $e_g$ states dictated by the cubic or
rhombohedral symmetry could be understood as effectively recovered by the superposition of two (or more) equivalent
ground states (of lower symmetry) having either of the $e_g$ orbitals occupied. To check this hypothesis and to obtain
one of these states, we have set the calculation in a larger unit cell of lower symmetry. 
This unit cell is described by the lattice vectors given by
${\bf v_1} = (-0.5, \, 0.5, \, 0), \, {\bf v_2} = (0, \, 1, \, -1), \, {\bf v_3} = (0.5, \, 0.5,\, 1)$
and contains 4 Cu and 4 O atoms. Each magnetic (111) plane contains two Cu atoms in this unit cell and they
are treated as of different kinds, albeit associated to the same pseudopotential. This artifact removes
the effective equivalence of $e_g$ states even for the 8 atoms cell description of the cubic structure.
A similar trick was also used for FeO to stabilize a broken symmetry (orbitally ordered) phase that reproduced
the structural distortions of the material under pressure~\cite{cococcioni2005linear}. The ground state
obtained in the 8 atoms cell has slightly lower energy per Cu-O pair ($\Delta E \simeq 1.88\, eV/{\rm CuO}$) 
compared to the rhombohedral 4 atoms unit cell, and thus the broken symmetry configuration is energetically favored.

\begin{figure}[!ht]
\includegraphics[width=0.45\textwidth]{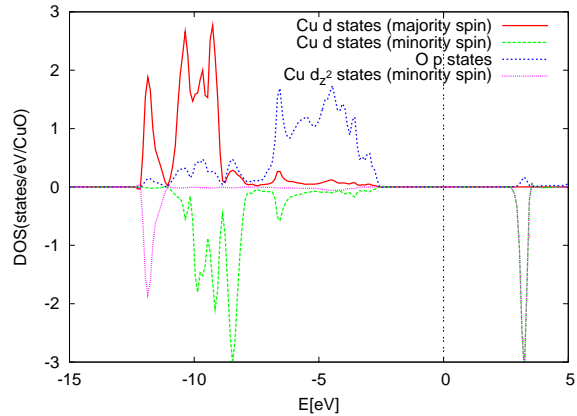}
\caption{\label{fig:cuo_gga_u_8} (Color online) The projected density of states in the broken symmetry phase.
                                 The on-site repulsion terms are $U_d = 9.79\, {\rm eV}$ and $U_p = 8.47\, {\rm eV}$
                                 (calculated from the response of GGA ground state).}
\end{figure}

It is important to remark that even in the broken symmetry phase, an energy gap appears only if a finite Hubbard
correction $U_p$ is used on the O p states. Without a Hubbard correction on O p states, the material is predicted
to be non-magnetic and a metallic ground state
still emerges from the degeneracy of the $e_g$ orbitals with opposite spin. This correction stabilizes the O p states
and increases their occupancy at the expense of lowering Cu d state occupancies. Thus, Cu d-orbitals are left with
9 electrons. Hund's rule magnetism favors the localization of the hole in this shell on one of 
the minority spin $e_g$ states.
The calculated d and p occupations reflect the localization of the hole:
$n^{\uparrow}_{\rm Cu}(e_g) \simeq 2.0$,  
$n^{\downarrow}_{\rm Cu}(d_{z^2}) \simeq 0.0$, $ n^{\downarrow}_{\rm Cu}(d_{x^2-y^2}) \simeq 1.0$,
while $n_{{\rm O}_p} = 5.51$.
These occupations also show that the Cu atoms acquire a finite magnetization which results in an AF-II ground state. 
The density of states of this ground state is shown in Fig.~\ref{fig:cuo_gga_u_8}.

Although the application of a Hubbard correction $U_p$ on non-correlated O p states is questionable, this computational
experiment is an indication of the fact that this system is characterized by a competition between two opposite 
tendencies: full occupation of Cu d states and the stabilization of a magnetic ground state through Hund's rule
coupling. If the number of electrons on d states is lower than a certain treshold value, then the Hund's rule 
magnetism is dominant, otherwise a non-magnetic ground state will appear. This competition is due to two factors:
a number of d electrons between 9 and 10 and O p states close in energy to the d states 
which are able to act as charge ``reservoirs" for them.
In the next section we further test this hypothesis by an extension to the +U corrective functional that 
explicitly includes a magnetic coupling J to encourage a magnetic ground state on each Cu atom.
%
\section{\label{sec:cubic_j}DFT+U+J functional and its application to the cubic phase}
%
The DFT+U functional introduced in equation (\ref{hub_u}) contains only a minimal set of on-site interaction parameters. 
In this section, we propose an extension of the DFT+U functional, that includes magnetic (exchange) 
interactions (DFT+U+J). While this is not new in literature (a review of previous approaches is given in reference
~\cite{ylvisaker2009anisotropy}), the functional we propose here deviates from previous formulations. The new 
corrective scheme can be obtained from a general second quantized expression for electron-electron interactions 
(derived in equation (6) of reference~\cite{jr2010extended}) given by
\begin{eqnarray}
\hat{V}_{\rm int} &=& \frac{1}{2} \sum_{I, \, J, \, K, \, L}\, \sum_{i, \, j, \, k, \, l}\,  \sum_{\sigma, \, \sigma'}\,
             \langle \phi_i^I \phi_j^J \vert V_{ee} \vert \phi_k^K \phi_l^L \rangle\, \nonumber\\
&& \qquad\qquad\qquad \times  \hat{c}_{I\, i\, \sigma}^{\dag}\, \hat{c}_{J \,j \, \sigma'}^{\dag}\, 
              \hat{c}_{K\, k\, \sigma'}\, \hat{c}_{L\, l\, \sigma} \label{Vint}
\end{eqnarray}
where capital letters $\{ I, \, \dots K\}$ represent site indices, lowercase letters $\{i, \, \dots k\}$ represent
state indices, $\{\sigma, \, \sigma'\}$ are spin indices; $V_{ee}$ denote the (screened) Coulomb interaction kernel
between electrons and $\phi_i^I$ denote the atomic wavefunction corresponding to state $i$ centered on site $I$. 
The operators $\hat{c}_{I\, i\, \sigma}^{\dag}, \, \hat{c}_{I, \, i\, \sigma}$ create/annihilate electrons with atomic
wavefunction $\phi_i^I$ and spin $\sigma$. Assuming that on-site interactions are dominant (especially for the localized
d states of transition-metal ions) we keep only terms with $I=J=K=L$ in the above sum. Moreover, we approximate 
the on-site effective interactions by the atomic averages of Coulomb and exchange terms:
$ U^I = \frac{1}{(2 l + 1)^2}\, \sum_{i, j} \langle \phi_i^I \phi_j^I \vert V_{ee} \vert \phi_j^I \phi_i^I \rangle $
and
$ J^I = \frac{1}{(2 l + 1)^2}\, \sum_{i, j} \langle \phi_i^I \phi_j^I \vert V_{ee} \vert \phi_i^I \phi_j^I \rangle $.
As a result, we obtain:
\begin{eqnarray}
{\rm E}_{\rm Hub} &=& \sum_{I, \, \sigma}\, \frac{U^I}{2}\, \left[ \left( n^{I\, \sigma} \right)^2 
                    + n^{I\, \sigma}\, n^{I\, -\sigma} - {\rm Tr}\left[ {\bf n}^{I\, \sigma}\, {\bf n}^{I\, \sigma} 
                    \right] \right]
                    \nonumber\\ 
              &&  \quad + \frac{J^I}{2}\, \left[ {\rm Tr}\left[ {\bf n}^{I\, \sigma}\, {\bf n}^{I\, \sigma} +
                    {\bf n}^{I\, \sigma}\, {\bf n}^{I\,-\sigma} \right] - \left( n^{I\, \sigma} \right)^2 \right]
\nonumber\\
                    \label{pj}
\end{eqnarray}
where the occupations 
$n^{I\, \sigma}_{i\, j} = \langle \hat{c}_{I\, i\, \sigma}^{\dag} \hat{c}_{I\, j\, \sigma} \rangle$  
are computed using the expression given in (\ref{occup});   
$n^{I\, \sigma} = {\rm Tr}[{\bf n}^{I\, \sigma}]$ and $n^I = \sum_{\sigma}\, n^{I\, \sigma}$. We introduce
a double counting term to be subtracted from $E_{\rm Hub}$ that is evaluated as the mean field approximation of 
(\ref{pj}) in the  fully localized limit~\cite{petukhov2003correlated}, where each atomic orbital is either filled 
by a single electron or totally empty. In this approximation we have:
\begin{equation}
{\rm Tr}[ {\bf n}^{I\, \sigma}\, {\bf n}^{I\, \sigma} ] \rightarrow n^{I\, \sigma} \,\,\, , \,\,\,
{\rm Tr}[ {\bf n}^{I\, \sigma}\, {\bf n}^{I\, -\sigma} ] \rightarrow n^{I\, \sigma_{\rm min}} \nonumber
\end{equation}
where $\sigma_{\rm min}$ denotes the minority spin. The above expression is true for both magnetic and 
non-magnetic systems (for non-magnetic systems $\sigma_{\rm min} = \sigma$, since spin up and down densities are
equivalent). In the fully localized limit, the entire double counting term thus reads
\begin{eqnarray}
E_{\rm dc} &=& \sum_{I}\, \frac{U^I}{2}\, n^I\, (n^I - 1) - 
             \sum_{I, \, \sigma}\, \frac{J^I}{2}\, n^{I\, \sigma}\, ( n^{I\, \sigma} -1 ) \nonumber\\
           && \qquad + \sum_I\, J^I\, n^{I\, \sigma_{\rm min}}. \label{dc}
\end{eqnarray}
The first term in the above equation is already included in the standard DFT+U functional given in equation (\ref{hub_u}).
After some algebra, we easily obtain the expression of the corrective functional as
\begin{eqnarray}
E_{\rm Hub} - E_{\rm dc} &=& \sum_{I, \, \sigma}\, \frac{U^I - J^I}{2}\, {\rm Tr}[ {\bf n}^{I\, \sigma}\, 
                                      ( {\bf 1} - {\bf n}^{I\, \sigma} ) ] \nonumber\\
            && + \sum_{I, \, \sigma}\, \frac{J^I}{2}\, 
                       \{ {\rm Tr}[ {\bf n}^{I\, \sigma}\, {\bf n}^{I\, -\sigma} ] -
                          2\, \delta^{\sigma\, \sigma_{\rm min}}\, n^{I\, \sigma} \}. \nonumber\\
                \label{dft_puj_1}
\end{eqnarray}
Comparing (\ref{hub_u}) and (\ref{dft_puj_1}), one can see that the on-site Coulomb repulsion parameter ($U^I$)
is effectively reduced
by $J^I$ for interactions between electrons of parallel spin and a positive $J$ term further discourages
anti-aligned spins on the same site. As a result, the functional given in equation (\ref{dft_puj_1}) 
encourages magnetic ordering.
Within the simple Dudarev model~\cite{dudarev1998electron}, the inclusion of $J$
has only been considered as the effective renormalization of U (i.e. $U^I \rightarrow U^I - J^I$) and the 
terms in the second line of (\ref{dft_puj_1}) were not included. The quadratic term in the second line
of equation (\ref{dft_puj_1}) can be explicitated as
\begin{equation}
\sum_{I, \, \sigma}\, \frac{J^I}{2}\, n^{I\, \sigma}_{m\, m'}\, n^{I\, -\sigma}_{m'\, m}. \label{explicitJ}
\end{equation}
Since the occupations can be understood as the expectation value 
$n^{I\, \sigma}_{m, \, m'}  = \langle \hat{c}_{I\, m\, \sigma}^{\dag}\, \hat{c}_{I\, m'\, \sigma} \rangle$, 
this term
describes an ``orbital exchange" between electrons of opposite spins (e.g. up spin electron from $m'$ to $m$
and down spin electron from $m$ to $m'$). It is important to notice that this term is genuinely beyond Hartree-Fock.
In fact, a single Slater determinant containing the four states $m \uparrow$ , $m \downarrow$, $m' \uparrow$ , 
$m' \downarrow$ would produce no interaction term like the one above. So this contribution to the corrective 
functional can be understood as resulting from the interactions between configurations that differ from each other
by two single electron states. In this context, the use of occupation numbers computed as in equation (\ref{occup})
is not legitimate (these configurations do not contribute together to any single term of the electronic charge 
density). Thus the expression of the $J$ term given in equation (\ref{dft_puj_1}), based on a product of 
${\bf n}^{I\, \sigma}$ and ${\bf n}^{I\, -\sigma}$ is an approximation of a functional that would require
the calculation of the 2-body density matrix.
Based on this reasoning, we argue that these interaction terms are not captured by approximate DFT
functionals, where the total energy is a functional of the one-body electron density. Therefore, we can suppose 
that they are completely missing from the DFT functional and we can neglect them in the double counting term 
that thus leads to
\begin{equation}
E_{\rm dc} = E_{\rm dc}^U - \sum_{I, \, \sigma}\, \frac{J^I}{2}\, n^{I\, \sigma}\, ( n^{I\, \sigma} - 1 )
\label{dc_1} 
\end{equation}
where $E_{\rm dc}^U = 1/2\, \sum_I\, U^I\, n^I\, ( n^1 -1 )$.
The double counting term (\ref{dc_1}) 
was previously considered in~\cite{anisimov1997first,czyżyk1994local}. It corresponds to the sum over like-spin 
electron pairs multiplied by the exchange parameter, and takes into account the total exchange energy in 
an average way. As a matter of fact, we have verified that that both dc terms (\ref{dc}) and (\ref{dc_1})
yield the same ground state for CuO. However, the one in equation (\ref{dc_1}) is numerically more stable and we have
adopted it in all calculations presented here.

Although never included in corrective DFT-based functionals, terms like in equation (\ref{explicitJ}) were introduced
in numerical studies based on model Hamiltonians~\cite{model1,model2}.

In order to calculate the Hubbard exchange parameter $J$, we have extended the linear response 
approach~\cite{cococcioni2005linear} used in the previous section and 
we have computed the responses of on-site magnetizations $m^J = n^{J\, \uparrow} - n^{J\, \downarrow}$
to a magnetic perturbation $\beta\, m^I$. Modeling the total energy of the solid with the double 
counting term (either equation (\ref{dc}) or (\ref{dc_1})), and rewriting it in terms of the on-site occupations $n^I$ and 
magnetizations $m^I$, we can calculate the exchange parameter $J^I$ from $\partial^2 E/(\partial m^I)^2 = -J^I/2$.
The second derivative of the energy with respect to on-site magnetizations are calculated using the response matrices
$\chi_{I\, J} = \partial m^I/\partial \beta^J$ so that $J^I = -2 [ (\chi^0)^{-1}_{II} - (\chi)^{-1}_{II} ]$. 
In this equation $\chi^0$ denotes the bare response matrix which is computed from the non-interacting Kohn-Sham 
problem, which needs to be subtracted from the response of the interacting system to obtain the value of $J^I$
as described in~\cite{cococcioni2005linear}. 

In this work, the $J$ parameter was computed using 32 atoms supercell and we found that $J \simeq 2.5\, eV$ (The 16 atoms
supercell employed for the calculation of U proved to be insufficient for obtaining linearly behaving magnetic 
response matrices). We would like to stress that
the values $U \simeq 9.79\, eV$ used in the previous section and $J \simeq 2.5\, eV$ are obtained 
by the response of the GGA ground state, and are used as ``test" values in the previous and current sections. More 
precise values are obtained by a self-consistent procedure (i.e. by recomputing the responses using the $GGA+U$
ground state) for the discussion of elongated structures in the next section.

\begin{figure}[!ht]
\includegraphics[width=0.45\textwidth]{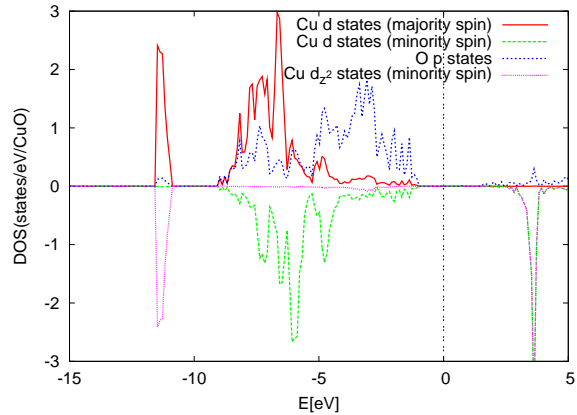}
\caption{\label{fig:cuo_gga_j_8} (Color online) The projected density of states in the broken symmetry phase.
                                 The Hubbard parameters for the Cu-d states are 
                                 $U = 9.79\, {\rm eV}$ and $J = 2.50\, {\rm eV}$
                                 (calculated from the response of GGA ground state).}
\end{figure}

In agreement with the discussion at the end of the previous section, the explicit account of magnetic interactions
through the new functional results in an insulating and antiferromagnetic ground state (with a broken symmetry 
phase). The resulting density of states is shown in Fig.~\ref{fig:cuo_gga_j_8}. The exchange interaction parameter
$J$ enhances the splitting between opposite spin electrons and favors a magnetic (insulating) state. 
As can be seen in Fig.~\ref{fig:cuo_gga_j_8}, the $GGA+U+J$ functional localizes a hole in the
$d_{z^2}$ state on each Cu atom, as for the case of the $GGA+U+U_p$ ground state, while all other d states are
filled an lie below the gap. This result suggests that the insulating ground state is stabilized by
magnetic interactions. 
Recently, the importance of the exchange coupling $J$ in favoring metallic or insulating ground states
of correlated systems has also been verified using the dynamical mean field 
theory~\cite{medici2011janus}.
However, magnetic and non-magnetic ground states are very close in energy. We hypothesize 
that this balance could be inverted by doping. We have also checked that it is possible to localize the hole on the
$d_{x^2-y^2}$ orbital or a configuration with mixed occupations (i.e. one hole localized on $d_{x^2-y^2}$ on one
Cu atom and one hole localized on $d_{z^2}$ on the other Cu atom of the same (111) plane). These configurations have
slightly higher energies than the ground state we have discussed above (the state with mixed occupations
is about $0.3\, eV/{\rm cell}$ higher in energy than the ground state, and the configuration with the $d_{x^2-y^2}$ hole
is about $0.5\, eV/{\rm cell}$ higher in energy than the ground state). 
The relatively low energy difference between them is due to the cubic crystal structure which is broken/lifted
on $e_g$ states for the electrons. 

As pointed out in the introduction, the broken symmetry insulating state in the cubic phase was never found before, 
and the degeneracy between the $e_g$ levels was lifted through a tetragonal distortion in other works
~\cite{grant2008electronic,peralta2009jahn,chen2009hybrid}. We have shown instead, that the symmetry can be broken even
for the cubic cell (with a lower symmetry 8 atoms unit cell, effectively corresponding to the cubic 
structure) and that an insulating state can result from magnetic interactios.
In the next section, we study elongated structures and determine their ground state properties using 
the 8 atoms cell.

\section{\label{sec:tetragonal}Tetragonally distorted structures}
\begin{figure}[!ht]
\includegraphics[width=0.45\textwidth]{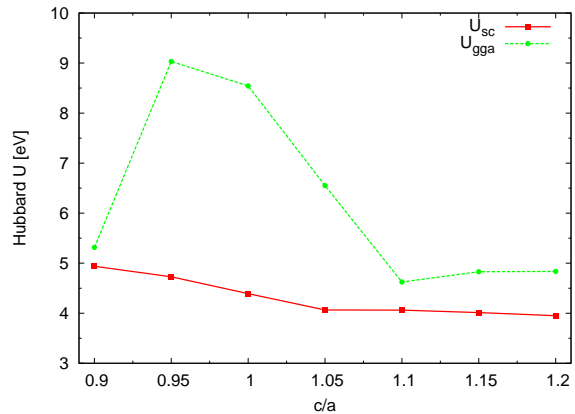}
\caption{\label{fig:u_vs_coa} (Color online) Calculated $U_d$ for each value of $c/a$.
                               The green line shows the linear response values calculated from the GGA 
                               response and the green line shows the self-consistently calculated values.}
\end{figure}

\begin{figure}[!ht]
\includegraphics[width=0.45\textwidth]{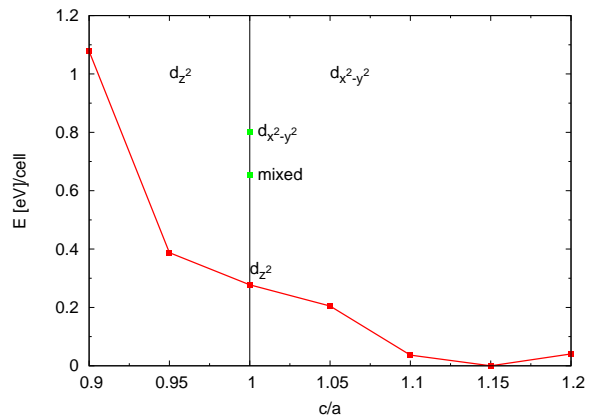}
\caption{\label{fig:e_vs_coa} (Color online) The ground state energy profile as a function of the tetragonal
                               distortion $c/a$. The orbital localizations of the holes on Cu d states for 
                               $c/a>1$ and for $c/a<1$ are labeled. The ground state energies of different hole
                               localizations for the cubic phase are also shown.}

\end{figure}

In this section we discuss the ground state properties of the tetragonally distorted structures. We limit
our study only to the case of AF-II ordering (unlike some previous studies~\cite{peralta2009jahn,chen2009hybrid},
which also considered other magnetic configurations) and determine the value of the tetragonal distortion
$c/a$ corresponding to lowest energy. To do so, we have calculated the Hubbard parameter $U$ at each value of $c/a$
between $0.9$ and $1.2$ using the linear response approach in a self-consistent procedure, while the $J$ parameter 
was fixed to the value obtained from the cubic cell and just with the GGA response (we assumed its variation
to be less important). In fact, the value of the parameter $J$
must be calculated from the response of a non-magnetic ground state (i.e. the GGA ground state of cubic phase
of CuO), since the linearity of the response matrices is not preserved when the ground state is magnetic
(i.e. GGA+U+J ground state, or any tetragonally distorted phase). 
Therefore, we have limited the calculation of $J$ to the non-magnetic phase. 
The $U$ parameters on the other hand, are computed self-consistently until their value converges within an 
accuracy of about $0.2\, eV$. The value of the lattice parameter
$a$ was fixed, so the volume of the cell varies between different calculations. However, we have also studied
a deformation at fixed volume and obtained very similar results, which will not be discussed in this work.
In Fig.~\ref{fig:u_vs_coa} we show the calculated values of Hubbard $U$ parameter as a function of $c/a$. We
show both the values calculated from GGA response (the green line) and the values that are calculated 
self-consistently (the red line). The self-consistent values of the $U$ parameters are smaller than the ones 
calculated from the GGA response, especially around the region close to $c/a \sim 1$ (i.e. the cubic phase). 
This difference is due to the fact that the GGA ground state in the cubic structure is metallic and paramagnetic,
while the $GGA+U+J$ ground state is insulating and antiferromagnetic. 
This effect is also visible at large tetragonal distortions, however it is less dramatic than for $c/a \simeq 1$,
since GGA yields ground states that are antiferromagnetic for $c/a \ga 1.1$ and $c/a \la 0.9$. From our calculations,
we find that the hole in the d states of Cu atoms are localized on the $d_{x^2-y^2}$ orbitals for $c/a > 1$
and on the $d_{z^2}$ orbitals for $c/a \le 1$. These orbital configurations are expected, since the elongation of 
the z-axis lowers the Coulomb repulsion energy of electrons localized on $d_{z^2}$ orbitals. Therefore,
the localization of the hole in the $d_{x^2-y^2}$ orbitals (or, equivalently, the localization of an electron on the 
$d_{z^2}$ orbitals) is energetically favorable for $c/a >1$ and vice-versa. The minimum energy configuration was
found to be at $c/a \simeq 1.15$ as shown in Fig.~\ref{fig:e_vs_coa}. The energy differences for different values
of $c/a$ are in overall agreement with the findings of previous studies~\cite{grant2008electronic,peralta2009jahn}.   
We have also calculated the energy band gaps for each structure, which lie between $1.4\, eV$ ($c/a=0.9$) 
and $0.4\, eV$ ($c/a=1.2$)
and decreases with $c/a$. The energy band gap for monoclinic CuO was determined to lie between 
$1.21\, eV$ and $1.7\, eV$ experimentally~\cite{koffyberg1982photoelectrochemical,marabelli1995optical}. 
The largest value of $1.4\, eV$ we have obtained is within the experimental range, but for larger values 
of $c/a$, the gap becomes lower than the experimental one. The difference is probably related with the fact that the 
structures we are considering have different symmetry than the ones studied experimentally. 

The value of the tetragonal distortion we found for the most stable configuration ($c/a \simeq 1.15$) 
is lower than the experimentally observed value of $c/a \simeq 1.35$. This difference could be related to the fact 
that our calculations do not take into account surface effects (strains) which are important for 
ultrathin films of tetragonal CuO grown on the SrTiO$_3$ support.
Indeed, it was recently shown that when surface effects are taken into account,
better agreement with experimental results are obtained~\cite{franchini2011thickness}.
The $c/a$ we found is in agreement with the results of 
references~\cite{grant2008electronic,peralta2009jahn}, however it is lower than
$c/a \simeq 1.377$ of reference~\cite{chen2009hybrid}. This difference could be related with the different 
localization properties of the hybrid-density functionals used in~\cite{chen2009hybrid} and DFT+U. The functional 
used in this work strongly localizes the electrons on atomic sites, and is less accurate in representing 
hybridization effects that could be important in CuO. The disagreement could 
be removed with the use of the inter-site interactions, which was shown to improve structural 
properties~\cite{jr2010extended}. In addition, a structurally consistent calculation of the Hubbard parameters 
as was done in~\cite{hsu2009first}, is expected to result in more precise structural properties.
Finally, we would like to stress that the local minimum located at $c/a \simeq 0.95$, which was identified
in some previous works~\cite{peralta2009jahn,chen2009hybrid}, has disappeared in our calculations, as can be seen in 
Fig.~\ref{fig:e_vs_coa}. Based on our results, we think that the local minimum was the consequence of the 
artificially high energy of the metallic cubic phase compared to the distorted ones. 
We argue that the metallic state obtained with the approximate DFT functional for $c/a=1$ results from the 
degeneracy of $e_g$ orbitals which is the result of cubic symmetry.

\section{\label{sec:summary}Summary}
In this work we have studied the electronic structure of CuO both in the cubic and tetragonal phases. We have 
identified the insulating state in the cubic structure, which was expected but never found before. 
The emergence of the cubic insulating state requires the breaking of symmetry in the electronic structure and 
leads to an orbitally ordered ground state. 
We have found that the insulating ground state results from a delicate balance between two tendencies:  
filling the d shell of Cu with (nearly) 10 electrons and localizing a hole on one of the $e_g$ states to 
stabilize a magnetic ground state.
After stabilizing the magnetic ground states, we have identified several local energy minima in the cubic 
configuration (paramagnetic, with holes localized on $d_{x^2-y^2}$ orbitals, on $d_{z^2}$ 
orbitals and with mixed type of localizations) at slightly higher energies.
We have also studied tetragonal distortions in the system and found
the lowest energy configuration to be at $c/a \simeq 1.15$. Our findings are in reasonable agreement with
experimental results, although inclusion of inter-site interactions in the functional could improve the agreement.
Finally, we clarified the transition (through the cubic phase with $c/a=1$) between the two different 
localization regimes of Cu d electrons ( on $d_{x^2-y^2}$ orbitals for $c/a \le 1$ and on $d_{z^2}$ orbitals for
$c/a > 1$) and suggested that the metallic state predicted by approximate DFT functional for the cubic phase
is the result of the degeneracy between $e_g$ states, artificially enforced by the symmetry of the crystal.
We believe that the interplay between orbital ordering and magnetism and the interaction between the d and p electrons, 
highlighted in this work, will be of interest in studying high T$_c$ superconductors, where similar 
electronic dynamics and competitions between charge and spin degrees of freedom are believed to play an important
role.

\begin{acknowledgments}
We acknowledge support from the NSF grant EAR-0810272, and the Minnesota Supercomputing Institute (MSI) for
providing computational resources.
We would also like to thank P. W. Grant for proposing the problem.
\end{acknowledgments}

\bibliography{hc_v5}

\end{document}